\documentclass[11pt]{article}

\usepackage{times}

\usepackage{amssymb}
\usepackage{mathtools}
\usepackage{graphicx}
\usepackage{rotating}
\usepackage{tikz}
\usetikzlibrary{arrows.meta}

\newenvironment{sciabstract}{%
\begin{quote} \bf}
{\end{quote}}

\newtheorem{theorem}{Theorem}
\newtheorem{corollary}{Corollary}
\newtheorem{proposition}{Proposition}

\newcommand*{\qed}{\null\nobreak\hfill\ensuremath{\square}}
\newcommand{\EE}{\mathbb{E}}
\newcommand{\var}{\mathrm{Var}}
\newcommand{\cov}{\mathrm{Cov}}
\newcommand{\sign}{\mathrm{sign}}
\newcommand{\N}{\mathrm{N}}
\newcommand{\indep}{\perp \!\!\! \perp}

\newcounter{lastnote}

\title{Risk Scores, Label Bias,\\and Everything but the Kitchen Sink} 

\author{
  Michael Zanger-Tishler\\
  Harvard University\\
  \and
  Julian Nyarko\\
  Stanford University\\
  \and
  Sharad Goel\\
  Harvard University\\
}

\date{}

\begin{document} 


\maketitle 


\begin{sciabstract}
  In designing risk assessment algorithms, many scholars promote a ``kitchen sink'' approach,
  reasoning that more information yields more accurate predictions.
 We show, however, that this rationale often fails when algorithms are trained to predict a proxy of the true outcome, as is typically the case.
 With such ``label bias'', one should exclude a feature if its correlation with the proxy and its correlation with the true outcome have opposite signs, conditional on the other model features. This criterion is often satisfied when a feature is weakly correlated with the true outcome, and, additionally, that feature and the true outcome are both direct causes of the remaining features. For example, due to patterns of police deployment, criminal behavior and geography may be weakly correlated and direct causes of one's criminal record, suggesting one should exclude geography in criminal risk assessments trained to predict arrest as a proxy for behavior.
\end{sciabstract}


\thispagestyle{empty}
\section{Introduction}

Risk assessments are central to the allocation of resources and the imposition of sanctions. 
In medicine, estimated health risks guide treatment decisions~\cite{mullainathan2022diagnosing}; in banking, default risk determines whether an applicant should be granted a loan~\cite{leo2019machine}; in education, the risk of non-completion is an important factor for college admissions decisions~\cite{aulck2019mining};
and in criminal justice, recidivism risk helps judges decide whether to detain or release a defendant while their cases proceed~\cite{chouldechova2017fair, dressel2018accuracy,lin2020limits}. Increasingly, the risk of such adverse events is estimated with the help of statistical algorithms. In training these algorithms, there is a widely shared view that the investigator should use as much data as is available to them \cite{berk2019machine,manski2022patient,manski2022using}. 
This view rests on the intuition that more information leads to (weakly) better predictions: If the added data are informative in estimating risk, then they will improve the performance of the algorithm, and if the added data do not contain a helpful signal, then they will be discarded without hurting performance. 
Proponents of this view stress that feature importance in the predictive context neither requires nor implies a causal link between algorithmic inputs and predicted outcomes~\cite{manski2022patient}. 
Absent the constraints of rigorous causal identification, it is argued that investigators can remain entirely atheoretical and simply hand all available data over to the predictive algorithm. 

In this paper, we show how ``label bias'', present in virtually all real-world scenarios in which algorithms are deployed today, can invalidate this common rationale. Label bias occurs when the outcome of interest is not observed directly, but is instead observed with measurement error. For instance, although criminal risk assessment tools seek to estimate the risk of future criminal \emph{behavior}, we typically only observe whether individuals are \emph{arrested} or \emph{convicted} of a crime. 
Similarly, tools to estimate health risk often seek to divert resources to the patients with the most significant medical \emph{needs}, but our observations are often limited to medical \emph{expenditures}. 
The inclusion of additional features will in general improve an algorithm's prediction of the proxy label (e.g., arrest or medical expenditures), but in the presence of label bias, the additional information can decrease the quality of predictions for the true label (e.g, criminal behavior or medical need). 
Below, we formally demonstrate and empirically illustrate conditions under which the inclusion of additional features hurts the predictive performance on the true outcome of interest. Because researchers rarely have access to the true label, whether or not to include a particular feature often rests on unverifiable assumptions about the relationships that gave rise to the proxy label. The findings highlight that most predictive contexts require investigators to spend significant time and care in developing a theoretical model of the underlying data generating process, thus removing one of the most important differentiators between prediction and causal inference.

Our study contributes to a burgeoning literature examining the use of algorithmic risk prediction in a variety of domains. These algorithms are frequently used to predict the risk of adverse events such as future criminal offending and failure to appear in court~\cite{imaiexperimental}, the risk of child abuse~\cite{brown2019toward,chouldechova2018case,shroff2017predictive}, money laundering~\cite{zhang2019machine}, students lagging behind in their learning \cite{cattell2021identifying}, and the risk of non-payment of loans~\cite{leo2019machine}. They are also used in situations where organizations or governments are deciding how to allocate scarce resources such as providing building permits \cite{mayer2013big}, assigning students to schools \cite{allman2022designing}, assigning high-risk patients to programs providing them more care \cite{obermeyer2019dissecting}, and determining who will receive kidney transplants \cite{friedewald2013kidney}. Further, corporations are currently using these tools to inform decisions about who receives information about housing advertisements~\cite{pmlr-v81-speicher18a} and employment opportunities~\cite{lambrecht2019algorithmic}. Algorithmic risk assessment tools can be better than humans at determining risk~\cite{goel2021accuracy}. However, scholars continue to critique these algorithms and study whether and under what conditions they can fairly and effectively be deployed in society~\cite{wang2022against,corbett2023measure,chouldechova2020snapshot,chohlas2023designing}.

In addition, our analysis builds on and contributes to a substantial body of literature examining the impact of label bias in statistical analyses. Prior work in the social sciences has long focused on the importance of measurement error for causal studies. Within this literature, a main focus has traditionally been on examining the importance of measurement error in the \textit{independent} variable, which can, at best, attenuate the causal estimates \cite[pp. 320--323]{wooldridge2015introductory}, and at worst, bias the coefficients in ways that are difficult to predict \cite{chalfin2018us}. Less attention has been given to label bias (i.e., measurement error in the dependent variable), perhaps because it is 
often assumed that proxy labels differ from the true labels by random noise, in which case one can still obtain unbiased causal estimates~\cite[pp. 318--320]{wooldridge2015introductory}. 
Existing research, however, suggests that there is a non-random relationship between the true and proxy labels across a variety of contexts, such as in the case of arrest and offending \cite{fogliato2021validity}.
More recent contributions have considered the impact of such systematic errors in the labels. For example, Knox et al.~\cite{knox2022testing} examine the potential for biases to arise in causal estimates when latent concepts that cannot be directly measured---like political ``ideology'' and ``democracy''---are approximated by proxy variables constructed from statistical models. 
Complementary work in computer science has examined the impact of label bias in a predictive setting. For instance, although predictive models may perform well on the proxy label, research has shown they are not guaranteed to be accurate on the true label if the measurement error between the true and proxy label is non-random
\cite{wang2021fair}. Similarly, label bias can also reduce the fairness of these algorithms on the true label \cite{fogliato2020fairness}. When feasible, training predictions on the true
label rather than a proxy has been shown to reduce racial inequalities in
algorithmic prediction and increase accuracy \cite{pierson2021algorithmic, mullainathan2021inequity, obermeyer2019dissecting}.

We build on these contributions by explicitly examining how the performance decrease from label bias interacts with the inclusion of additional predictors into the model. 
To establish our results, we begin, in Section~\ref{sec:stats}, by deriving analytic conditions for when excluding factors in a model trained to predict a proxy label is guaranteed to improve predictions of the true outcome of interest.
We demonstrate and build intuition for these analytic results using a stylized example of estimating recidivism risk in the presence of label bias,
where reoffense is the true label of interest and rearrest is the observed proxy.  
Then, in Section~\ref{sec:applications}, 
we turn to two case studies.
First, we consider partially synthetic recidivism data with real rearrest outcomes (the proxy label) and simulated  
reoffense outcomes (the true label).
This setting resembles one that many researchers face in practice, where data on the true label are often prohibitively difficult or impossible to obtain. 
We show how different assumptions about how the true label relates to the observed proxy affect decisions about what predictors to include in the risk assessment model. 
Second, we consider a dataset from the health sciences. In targeting patients for high-risk care management programs, we rely on data by Obermeyer et al.~\cite{obermeyer2019dissecting} which contain, among other items, information on both the true label (healthcare need) and a proxy (healthcare spending). Using this dataset, we estimate the 
welfare costs of using a kitchen-sink predictive model instead of more judiciously selecting a model that accounts for label bias.
We conclude in Section~\ref{sec:conclusion} with a discussion of our findings and point out potential paths forward.

\section{A Statistical Condition for Excluding Features}
\label{sec:stats}

To build intuition for how label bias impacts the choice of features in predictive models, we start with a simplified motivating example from the criminal justice context.
In the United States, after an arrest, a judge will often decide whether or not to detain the arrested individual based on their estimated risk to public safety.
In practice, this risk is commonly estimated using statistical risk assessments. The underlying risk models are trained using information about future arrests and convictions. However, arrests and convictions are not direct measures of public safety risks. Instead, they merely act as proxies, making these risk assessment tools susceptible to label bias.

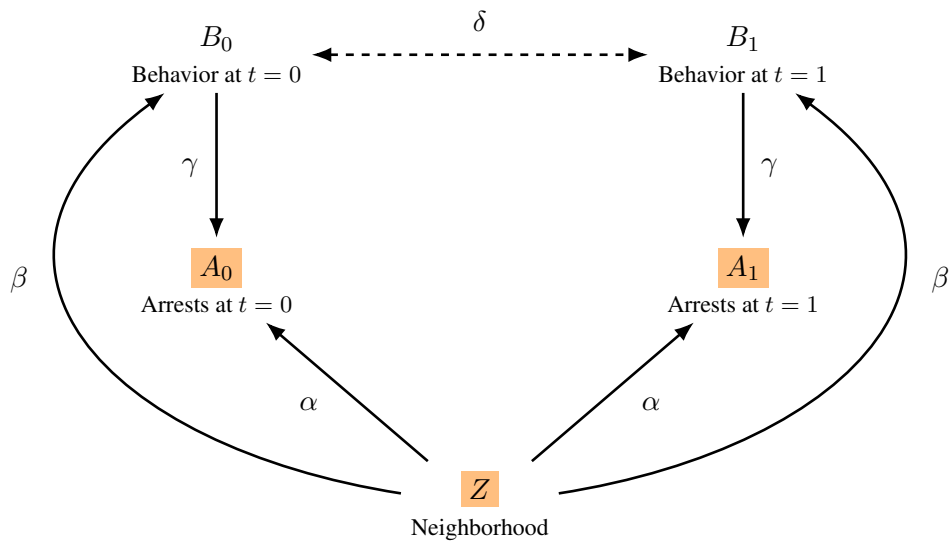
\begin{figure}[t]
  \vspace{5mm}
    \begin{center}
    \begin{tikzpicture}[xscale = 3.5, yscale = 3, align = center]
      \node (Z)      at (0, 0)
        {\colorbox{orange!50}{\(Z\)}\\{\footnotesize Neighborhood}};
      \node (A0)      at (-1, 1)
        {\colorbox{orange!50}{\(A_0\)}\\{\footnotesize Arrests at \(t=0\)}};
      \node (A1)      at (1, 1)
        {\colorbox{orange!50}{\(A_1\)}\\{\footnotesize Arrests at \(t=1\)}};  
      \node (B0)      at (-1, 2)
        {\(B_0\)\\{\footnotesize Behavior at \(t=0\)}};
      \node (B1)      at (1, 2)
        {\(B_1\)\\{\footnotesize Behavior at \(t=1\)}};

      \draw[-Latex, line width=1pt]      (Z) to (A0);
      \draw[-Latex, line width=1pt]      (Z) to (A1);
      \draw[-Latex, line width=1pt]      (B0) to (A0);
      \draw[-Latex, line width=1pt]      (B1) to (A1);
      \draw[Latex-Latex, line width=1pt, dashed]      (B0) to (B1);
      \draw[-Latex, line width=1pt]      (Z) to [out=170, in=220, looseness=1.5] (B0);
      \draw[-Latex, line width=1pt]      (Z) to [out=10, in=320, looseness=1.5] (B1);

      \node at (-.65, .45) {\(\alpha\)};
      \node at (.65, .45) {\(\alpha\)};
      \node at (-1.75, 1) {\(\beta\)};
      \node at (1.75, 1) {\(\beta\)};
      \node at (-1.1, 1.5) {\(\gamma\)};
      \node at (1.1, 1.5) {\(\gamma\)};
      \node at (0, 2.15) {\(\delta\)};
    \end{tikzpicture}
  \end{center}
  \caption{\emph{%
      The data-generating process for our stylized example of criminal behavior (true label) and arrest (proxy label), with observed variables highlighted in orange. 
  }}
\label{fig:sim-dag}
\end{figure}

In Figure~\ref{fig:sim-dag}, we sketch the data-generating process for a stylized, linear structural equation model (SEM)~\cite{pearl2013linear} of arrests and behavior, where we treat arrests as the observed proxy for unobserved behavior, our true outcome of interest.
The model produces synthetic data on individual-level behavior (\(B_0\) and \(B_1\)) and arrest (\(A_0\) and \(A_1\)) outcomes at two time periods (\(t=0\) and \(t=1\)), as well as the neighborhood (\(Z\)) in which the individual resides. 
Importantly, arrests depend both on behavior and on neighborhood, reflecting the fact that people who engage in the same behavior may be arrested at different rates depending on where they live.
For example, Beckett et al.~\cite{beckett2006race}
found that the geographic concentration of police resources in Seattle led to higher arrest rates for Black individuals delivering drugs compared to white individuals delivering drugs---where the true racial distribution of those delivering drugs was estimated from survey data and ethnographic observations.
Similarly, Cai et al.\ \cite{cai2022measuring} found that the issuance of speeding tickets varied across neighborhoods even after adjusting for the true, underlying incidence of speeding, as estimated by the movement of mobile phones.

In this SEM, all of the variables are normally distributed, with mean 0 and variance 1. 
We can thus interpret their values as representing the extent to which individuals differ from the population averages.
In the case of neighborhood (\(Z\)), we can think of its value as denoting the level of police enforcement in that area.
Further details about the model are provided in the Appendix.

\begin{figure}[t]
\centering
\includegraphics[width=6in]{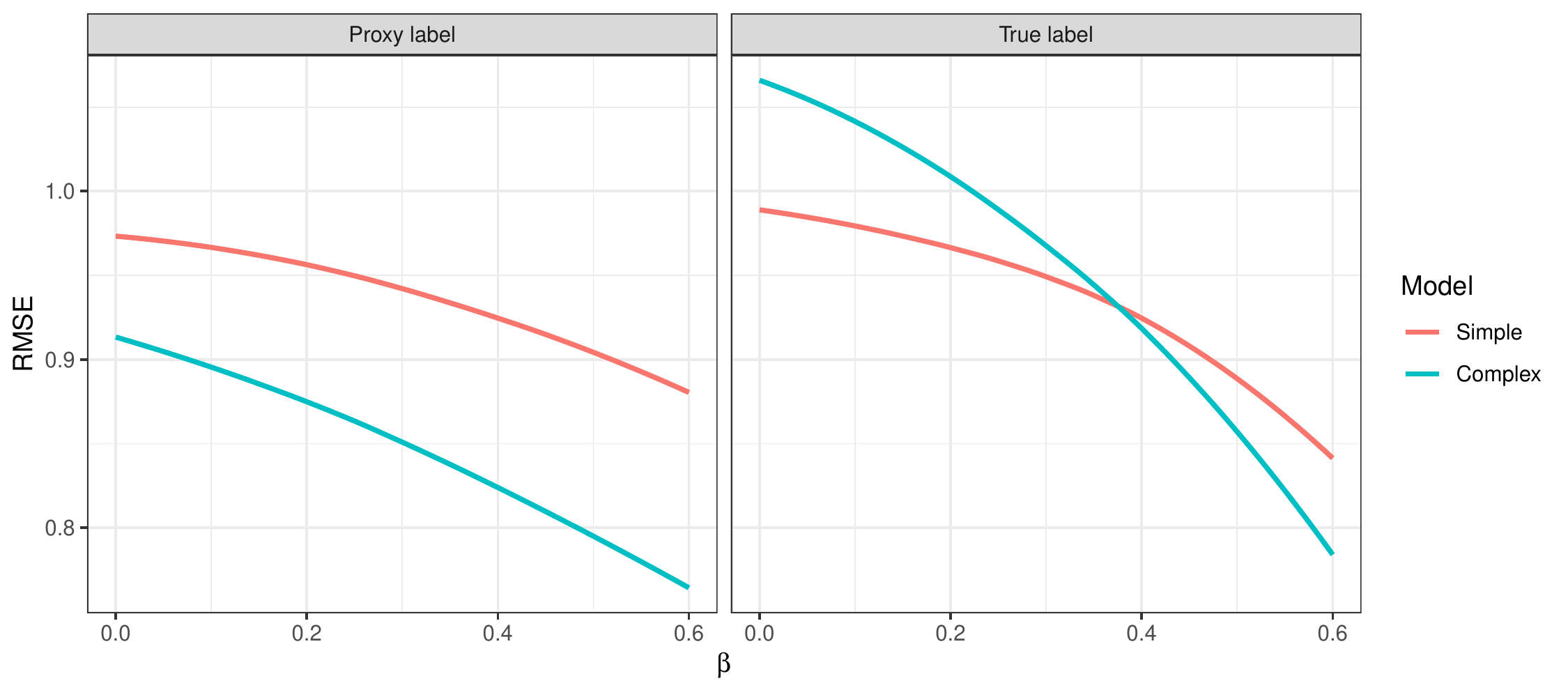}
\caption{\emph{Performance of simple and complex models trained to predict a proxy label, when evaluated on the proxy label (left) and the true label (right) for a range of \(\beta\) values.
Whereas the complex model outperforms the simple model on the proxy label, the simple model outperforms the complex model on the true label for certain values of \(\beta\).
}}
\label{fig:full-sim}
\end{figure}

Using synthetic data generated with this SEM, we train a ``complex'', kitchen-sink model to predict arrests at time \(t=1\) (\(A_1\)) based on arrests at time \(t=0\) (\(A_0\)) and neighborhood (\(Z\)).
The more parsimonious, ``simple'' model bases its predictions only on arrests at time \(t=0\), omitting neighborhood.
We now examine how the performance of the complex and simple models vary for different values of $\beta$, the parameter that describes the relationship between neighborhood and behavior, holding the other parameters fixed.\footnote{%
For this simulation, we set \(\alpha = \gamma = \delta = 0.4\), though the general pattern is largely invariant to this choice, as we describe in more detail below.
}
Across values of \(\beta\), the left-hand panel of Figure~\ref{fig:full-sim}
shows that the complex model outperforms the simple model---in terms of root mean squared error 
(RMSE)---when evaluated on the proxy label.
As expected, including more information reduces error when evaluated on the label used to train the models, a pattern that has traditionally motivated the inclusion of more features in predictive models.
However, moving to the right-hand panel of Figure~\ref{fig:full-sim}, 
we see that the simple model outperforms the complex model 
on the \emph{true} label for some values of \(\beta\).
In particular, the simple model outperforms the complex one for small values of \(\beta\), corresponding to a weak 
relationship between neighborhood and behavior.

Our SEM illustrates a scenario in which simple models outperform more complex models due to the presence of label bias. To understand this result, imagine two individuals, both of whom have the same prior arrest record, but with only one of them living in a heavily policed neighborhood. Further assume that where one lives has little impact on criminal behavior (corresponding to small \(\beta\)), but that heavier policing increases the chance of being arrested for an offense. In this case, we can infer that the individual living in the heavily policed neighborhood 
engaged in past criminal activity less frequently than
the individual living in the less heavily policed neighborhood. This is because fewer actual offenses are required to build a given arrest record in areas of high enforcement.
Extrapolating from their past behavior, we would accordingly expect the individual in the heavily policed area to be less likely to engage in future criminal behavior.
Thus, using information about one's neighborhood to predict future arrests (the proxy label) correctly tells us that the individual living in the heavily policed neighborhood is more likely to be rearrested,
but it incorrectly suggests that individual is also more likely to engage in future criminal behavior (the true label).
So, when predicting arrests as a proxy for behavior, it is better in this case to exclude information on one's neighborhood.

The SEM depicts a specific data-generating process, but the phenomenon we identify is generalizable.
Theorem~\ref{thm:main} and Corrolary~\ref{cor:main} below establish formal conditions under which this pattern is guaranteed to occur.

\begin{theorem}
\label{thm:main}
Suppose \(Y\) and \(Y'\) are two arbitrary random variables, where \(Y\) is the ``true'' outcome of interest and \(Y'\) is a proxy. For a random variable \(Z\) and a vector of random variables \(X = (X_1, \dots, X_k)\), consider the estimators
\begin{align*}
\hat{Y}_{X,Z} &= \EE[Y' \mid X, Z], \ \text{and}\\
\hat{Y}_X &= \EE[Y' \mid X],
\end{align*}
where \(\hat{Y}_{X,Z}\) is the ``complex'' estimator that uses all available features, and \(\hat{Y}_X\) is the ``simple'' estimator that omits \(Z\).  
If  
\(\cov\left(\hat{Y}_{X,Z}, Y \mid X\right) \leq 0\),
then \(\EE\left[\left(\hat{Y}_X - Y\right)^2\right] \leq \EE\left[\left(\hat{Y}_{X,Z} - Y\right)^2\right]\), meaning the simple estimator weakly outperforms the complex estimator.
Further, if \(\EE\left[\var\left(\hat{Y}_{X,Z} \mid X\right)\right] \neq 0\), 
then the simple estimator strictly outperforms the complex estimator.
\end{theorem}

In the setting of Theorem~\ref{thm:main}, one seeks to estimate a true outcome of interest \(Y\), and must choose between two different estimators designed to predict the proxy label \(Y'\). 
The first, ``complex'' estimator (\(\hat{Y}_{X,Z}\))
uses both \(X\) and \(Z\) to predict \(Y'\),
whereas the second (\(\hat{Y}_{X}\)) uses only \(X\).
The theorem shows that if, conditional on \(X\), the
true label (\(Y\)) is negatively correlated with the complex estimator (\(\hat{Y}_{X,Z}\)), then
the simple model generally outperforms the complex estimator on the true outcome of interest.
If, alternatively, the true and proxy labels differ only by additive, independent noise, then Proposition~\ref{prop:noise} in the Appendix shows that including more information when predicting the proxy label will in general improve predictive performance on the true label.
Thus, in the absence of systematic measurement error---including the case where there is no measurement error---that result confirms the conventional wisdom that more information is better .

When the complex estimator \(\hat{Y}_{X,Z}\) is 
linear in \(Z\), 
Corollary~\ref{cor:main} establishes a simpler condition under which performance increases by omitting information.
Specifically, if, conditional on \(X\), 
\(Z\) is positively correlated with true label \(Y\) but negatively correlated with the proxy label \(Y'\) (or vice versa), then omitting \(Z\) when predicting the proxy label will in general improve performance on the true outcome of interest.

\begin{corollary}
\label{cor:main}
In the setting of Theorem~\ref{thm:main}, suppose \(\hat{Y}_{X,Z}\) is linear in \(Z\), i.e., \(\hat{Y}_{X,Z} = f(X) + g(X) \cdot Z\) for some functions \(f\) and \(g\).
If 
\[\sign\left(\cov\left(Y, Z \mid X\right)\right) = -\sign\left(\cov\left(Y', Z \mid X\right)\right),\]
then \(\EE\left[\left(\hat{Y}_X - Y\right)^2\right] \leq \EE\left[\left(\hat{Y}_{X,Z} - Y\right)^2\right]\).
Further, if \(\EE\left[\var\left(\hat{Y}_{X,Z} \mid X\right)\right] \neq 0\), 
then the simple estimator strictly outperforms the complex estimator.
\end{corollary}

The linearity assumption of Corollary~\ref{cor:main} holds in a variety of settings. In particular, as described in the Appendix, it holds when \(Y'\), \(X\), and \(Z\) are jointly multivariate normal, as is the case in our SEM above.
To apply the corollary, one needs information on the correlations of \(Y\) and \(Z\) and of \(Y'\) and \(Z\), conditional on \(X\).
The former involves directly observed quantities---the proxy label and the potential predictors---and so, in practice, can be computed from data. For our stylized SEM, we show in the Appendix that this correlation is positive for all (non-degenerate) parameter choices, meaning that neighborhood (\(Z\)) is positively correlated with future arrests (\(A_1\)), conditional on past arrests (\(A_0\)).
The second conditional correlation we must consider when applying Corollary~\ref{cor:main}---the correlation between \(Y\) and \(Z\), conditional on \(X\)---is not typically directly observed, as it depends on the true label \(Y\).
Understanding its sign thus involves assumptions about how the true label is related to the predictors \(Z\) and \(X\).
For our SEM, we show in the Appendix that this correlation is negative for small values of \(\beta\).
That is, when \(\beta\) is small, 
neighborhood (\(Z\)) and future behavior (\(B_1\))
are negatively correlated conditional on past arrests (\(A_0\)).
Intuitively, this is because \(A_0\) is a collider, 
and so when we fix its value, increasing \(Z\) requires 
decreasing \(B_0\), which in turn decreases \(B_1\).
Thus, for small values of \(\beta\), omitting neighborhood when predicting the proxy label improves performance on the true label, as shown in Figure~\ref{fig:arrest-sim}.

\section{Case Studies}
\label{sec:applications}

To better understand the practical implications of our results, we now turn to two real-world datasets. 
The first allows us to further consider criminal risk assessments, adding additional realism to our stylized SEM above;
the second dataset comes from the medical domain, where the goal of the risk assessment we consider is to identify patients with complex healthcare needs.

\subsection{Criminal risk assessments}

Continuing with our running example studying arrest and criminal behavior, we use data on individuals from a major U.S. county who were arrested for a felony offense between 2013 and 2019. For simplicity, we limit the sample to the 25,918 cases where the individuals’ race was identified as either Black or non-Hispanic white. The dataset includes further details on each case, 
including information on the charges, the location, date and time of the incident, and the criminal history of the arrested individual. 
In addition, the dataset contains information on future rearrests, which we use as our proxy label for future offenses. Using these data, we fit simple and complex models trained on the proxy label (future arrests). We then examine model performance on the true label (future criminal offenses, which we simulate, as described below, since they are not directly observed). 
Our ``complex'' model includes 
three features: 
the age of the arrested individual;
the number of times the individual was previously arrested;
and whether or not the arrest occurred in a ``high policing'' area (i.e., a police district accounting for disproportionately high numbers of arrests).
Our ``simple'' model includes age and number of past arrests, but not location information---similar to many commonly used criminal risk assessment tools.

This example mirrors many instances of label bias in the real world, as it 
is 
difficult, and perhaps impossible, to directly estimate the risk of ``true'' offending~\cite{biderman1967exploring}.
This is in part because criminal behavior that is not reported to the police will not be included in administrative records. 
We thus simulate offending outcomes under a range of data-generating processes, and then examine how assumptions about criminal behavior affect model performance after including or omitting location information.
In particular, for a fixed value of \(\rho \in \mathbb{R}\),
describing the impact of neighborhood on criminal behavior,
we assume that each individual in our dataset commits a future offense with the following probability:

\begin{equation*}
    \label{eq:simulation-arrest-1}
   \Pr(B_1 = 1) = \text{logit}^{-1}\left(-1 - \frac{1}{100} X_{\text{age}} + \frac{1}{2}A_0  + \rho Z\right),
\end{equation*}
where \(B_1\) indicates future criminal behavior (our true label),
\(X_{\text{age}}\) is the arrested individual's age,
\(A_0\) is the number of times they were previously arrested,
and \(Z\) indicates whether the arrest took place in a high-policing area.
The intercept and the coefficients for 
\(A_0\) and \(X_{\text{age}}\)
were selected to approximate the coefficients from a regression of future arrests on age and past arrests in our data.

\begin{figure}[t]
\centering
\includegraphics[width=4in]{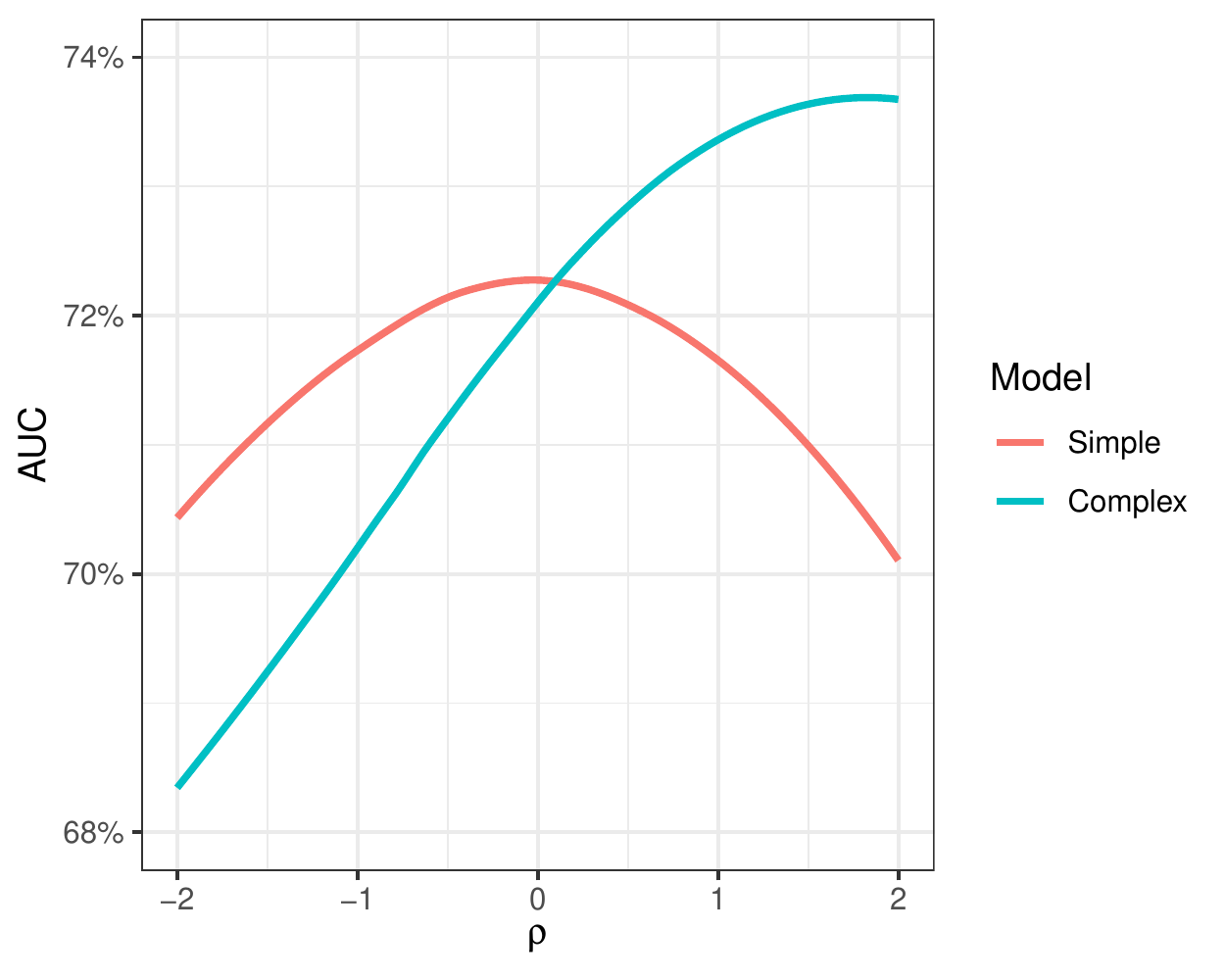}
\caption{\emph{Performance of simple (age and past arrests) and complex (age, past arrests, and neighborhood) models trained to predict future arrests (the proxy label), evaluated on
future criminal behavior (the true label).
Because the future criminal behavior is not directly observable, 
the plot shows results for synthetic outcomes generated under a range of data-generating processes parameterized by \(\rho\), the hypothesized relationship between neighborhood and future criminal behavior.}}
\label{fig:arrest-sim}
\end{figure}

Based on the data-generating process described above,
we now evaluate the ability of our simple and complex risk assessment models to predict the synthetic true label, future criminal behavior.
We evaluate model performance in terms of AUC, as the outcome is binary.%
\footnote{%
AUC is a common measure of performance in the machine learning community when considering binary outcomes.
Given a random individual who engaged in future criminal activity and a random individual who did not, the AUC of a risk assessment model is the probability that the model correctly identifies the individual in the pair who engaged in criminal activity. 
Our formal theoretical results are stated in terms of RMSE,
but this example and our subsequent example show that the general pattern and intuition extend to other popular evaluation metrics.} 
Figure~\ref{fig:arrest-sim}
shows that
the simple model outperforms the complex model on the true label when $\rho$ is negative, and the complex model outperforms the simple model when $\rho$ is sufficiently positive. 
Given two arrested individuals who are the same age and have the same number of past arrests, 
negative values of $\rho$ indicate that the individual who was arrested in the high-policing area is the less likely of the pair to engage in future criminal behavior.
That pattern is akin to what we saw in our stylized SEM depicted in Figure~\ref{fig:sim-dag}.
Accordingly, to the extent that one believes the hypothesized data-generating process with negative $\rho$
is a sufficiently accurate description of criminal behavior,
it is better to exclude neighborhood information when training risk assessment tools on the proxy label, future arrests.

\subsection{Identifying high-needs patients}
We continue by applying our results to a well-known case of label bias in the literature, that of a commercial risk assessment tool that health systems rely on to target patients for ``high-risk care management'' programs \cite{obermeyer2019dissecting}. 
These programs seek to 
enroll patients with complex medical needs, and subsequently provide them with a higher level of care. 
Because these programs are capacity constrained, the role of statistical risk assessments in this case is to 
accurately identify patients who would benefit the most from the additional care.
In practice, though, 
the risk assessment algorithms are often designed to predict future medical expenditures, 
a proxy for medical need as the true outcome of interest.
Analyzing these algorithms, Obermeyer et al.~\cite{obermeyer2019dissecting} 
conclude that, due to label bias, Black patients are less likely to be enrolled in the program than white patients with the same level of medical need. This is because unequal access to healthcare means that white individuals are more likely to seek medical treatment---and accordingly incur higher medical costs---than equally sick racial minorities.

Obermeyer et al.~\cite{obermeyer2019dissecting} highlight the importance of appropriately selecting the target of prediction,
and illustrate the accuracy and equity gains one can achieve by switching from predicting expenditures to a more direct measure of medical need.
Here we revisit the problem, and investigate how the choice of risk factors used to identify patients impacts enrollment decisions.
To do so, we start with the data released by Obermeyer et al.~\cite{obermeyer2019dissecting}, 
which include detailed information on
patient demographics (sex, race, and age),
current and future health, 
and past and future medical expenditures.\footnote{Obermeyer et al.~\cite{obermeyer2019dissecting} released a synthetic dataset, with variables having the same conditional distributions as those in the original dataset, using the \texttt{synthpop} package in R (the original data cannot be released in order to protect patient privacy). The data are available at: https://gitlab.com/labsysmed/dissecting-bias).} 
We then train simple and complex models on the proxy label, future medical costs. 
Our complex model includes all information available at the time of the enrollment decision (i.e., patient demographics, current health, and past medical expenditures);
our simple model includes only current health, excluding past medical medical costs and demographic variables. 
In the end, the complex model includes 150 predictors, and the simple model includes 128 predictors. 

\begin{figure}[t]
\centering
\includegraphics[height=2.75in,trim={0 0 1in 0},clip]{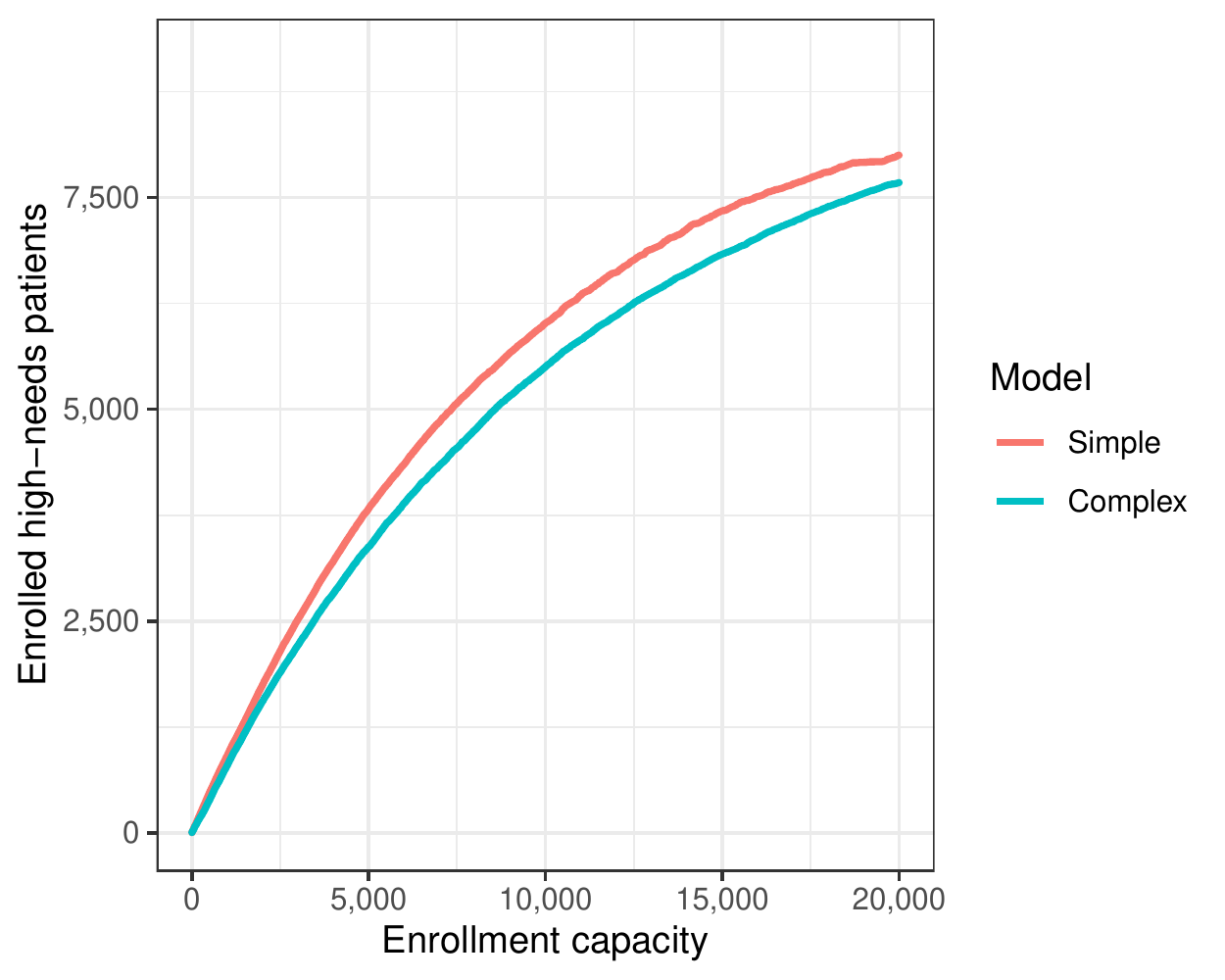}
\includegraphics[height=2.75in]{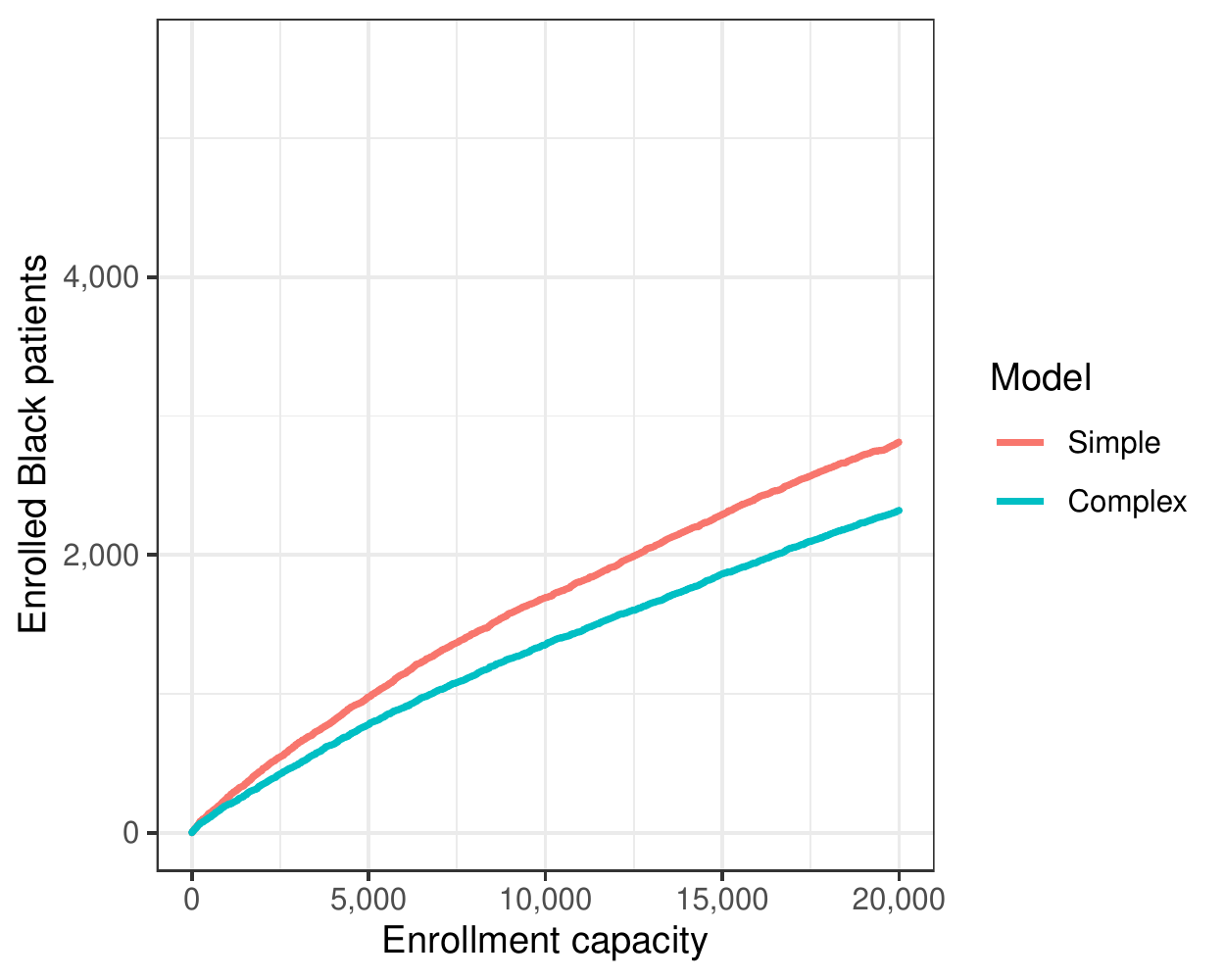}
\caption{\emph{Enrollment of high needs patients (left) and demographic composition of enrolled patients (right) under the simple and complex models for a range of program capacities.}}
\label{fig:obermeyer_enrollment}
\end{figure}

Finally, we evaluate both models on their ability to predict whether a patient, in the subsequent year, is found to suffer from at least three chronic diseases---a measure of future health need identified by Obermeyer et al.~\cite{obermeyer2019dissecting}.
The left-hand panel of Figure~\ref{fig:obermeyer_enrollment} shows the number of high-needs patients enrolled under the simple and complex models at different enrollment capacities, where the patients with highest estimated risk under the respective models are enrolled in the program.
At each capacity level, the simple model outperforms the complex model in identifying more high-needs patients. 
Additionally, as shown in the right-hand panel of Figure~\ref{fig:obermeyer_enrollment},
the simple model enrolls more Black patients than the complex model at every capacity level. 
This pattern stems from the simple model prioritizing patients with high expected medical needs over patients with high expected medical expenditures---the latter population being disproportionately white.
Thus, if one only has access to a proxy label, systematically excluding predictors in a risk assessment tool can improve both the accuracy and equity of the instrument.

\section{Conclusion}
\label{sec:conclusion}
In building predictive models, the traditional guidance is to include all available information to maximize performance.
But, as we have shown, a more judicious selection of features can lead to better model performance in the presence of label bias.
Because the true label of interest is often not readily available, it raises the question of what examiners should and can do to mitigate the negative consequences from taking a kitchen-sink approach to prediction. 
The examples we have discussed highlight several approaches that vary in their appropriateness based on data availability and understanding of the underlying data-generating process.

Most directly, Obermeyer et al.~\cite{obermeyer2019dissecting} illustrate how some instances of label bias can be addressed simply by making a more concentrated effort to collect data on the true label of interest. 
If such an effort is generally possible, but prohibitively costly, investigators should consider whether the true label of interest can be obtained for a smaller subset of the population. This subset, even if it is not sufficiently large to train models predicting the true label, might still be used to explore how the selection of features affects model performance on the true label.
If obtaining the true label is impossible, but investigators have access to a wealth of other features, one may simulate the true label of interest. In doing so, researchers should use their domain-specific knowledge to make reasonable assumptions about the relationship between the true label of interest and the features in question. We illustrated this process using felony offense data. Importantly, investigators need not constrain themselves to one particular relationship between the true label and the features, but can instead assess the sensitivity of feature selection to label bias across a wide range of plausible assumptions. 
Finally, investigators can make additional theoretical assumptions about the data-generating process in order to determine how label bias affects the choice of risk factors in a specific application---as we did in our healthcare example. As shown in that example, caution is particularly warranted for features that do not appear to be directly risk relevant. These features often yield little improvement on the true outcome of interest, and raise the likelihood that performance may decrease or that their inclusion may exacerbate disparities.

More generally, our findings suggest, in contrast to conventional wisdom, that one cannot entirely divorce the predictive enterprise from theoretical considerations. Instead, a successful deployment of predictive tools often rests on the plausibility of the assumptions about the underlying processes that give rise to the observed data, highlighting the continued utility of domain-specific expertise in the predictive context.

\bibliography{scibib}
\bibliographystyle{unsrt}

\section*{Acknowledgements}
We thank Avi Feller, Johann Gaebler, Talia Gillis, and Josh Grossman for helpful feedback and conversations.

\clearpage
\appendix

\section{Proof of Theorem~\ref{thm:main}}

For any random variable \(\hat{Y}\), 
\begin{equation*}
\EE\left[\left(\hat{Y} - Y\right)^2\right] = \EE\left[Y^2\right] + \EE\left[\hat{Y}^2\right] - 2\EE\left[Y\cdot\hat{Y}\right],
\end{equation*}
and so,
\begin{align}
\label{eq:diff}
\begin{split}
&\EE\left[\left(\hat{Y}_{X,Z} - Y\right)^2\right] - \EE\left[\left(\hat{Y}_X - Y\right)^2\right] \\
&\hspace{1in}= \left(\EE\left[\hat{Y}_{X,Z}^2\right] - \EE\left[\hat{Y}_X^2\right]\right) + 2\Big(\EE\left[Y\cdot\hat{Y}_X\right] - \EE\left[Y\cdot\hat{Y}_{X,K}\right]\Big).
\end{split}
\end{align}
We will show, in turn, that each of the summands on right-hand side of Eq.~\eqref{eq:diff} are non-negative---and that the former is strictly positive when \(\EE\left[\var\left(\hat{Y}_{X,Z} \mid X\right)\right] \neq 0\). To start, we note that
\begin{align}
\label{eq:jensen}
\begin{split}
\EE\left[\hat{Y}_{X,Z}^2 \mid X\right] &\geq \left(\EE\left[\hat{Y}_{X,Z} \mid X \right]\right)^2 \\
&= \left(\EE\left[\EE[Y' \mid X,Z] \mid X\right]\right)^2\\
&= \left(\EE\left[Y' \mid X\right]\right)^2\\
&= \hat{Y}_X^2,
\end{split}
\end{align}
where the first line follows from Jensen's inequality, 
and the second equality follows from the law of iterated expectations. 
As a result, by another application of the law of iterated expectations,
\begin{align}
\label{eq:ineq}
\begin{split}
\EE\left[\hat{Y}_{X,Z}^2\right] &= \EE\left[\EE\left[\hat{Y}_{X,Z}^2 \mid X\right]\right]\\
& \geq \EE\left[\hat{Y}_X^2\right],
\end{split}
\end{align}
showing that the first summand on right-hand side of Eq.~\eqref{eq:diff} is non-negative.
Further, the inequality in Eq.~\eqref{eq:jensen} is strict on the set 
where \(\var\left(\hat{Y}_{X,Z} \mid X\right) \neq 0\).
Consequently, if \(\EE\left[\var\left(\hat{Y}_{X,Z} \mid X\right)\right] \neq 0\), then 
\(\var\left(\hat{Y}_{X,Z} \mid X\right) \neq 0\) 
on a set of positive measure, 
and so the inequality Eq.~\eqref{eq:ineq} is likewise strict.

Now, turning to the second summand, we have
\begin{align*}
\EE\left[Y\cdot\hat{Y}_{X,Z} \right] &= \EE\left[\EE\left[Y\cdot\hat{Y}_{X,Z} \mid X\right] \right] \\
&\leq \EE\left[\EE\left[\hat{Y}_{X,Z} \mid X\right] \cdot \EE\left[Y \mid X\right]\right] \\
&= \EE\left[\hat{Y}_X \cdot \EE\left[Y \mid X\right]\right] \\
&= \EE\left[\EE\left[Y\cdot\hat{Y}_X \mid X\right]\right]\\
&= \EE\left[Y\cdot\hat{Y}_X\right], \\
\end{align*}
where we repeatedly applied the law of iterated expectations, used the fact that \(\hat{Y}_X\) is measurable with respect to \(X\) to establish the penultimate equality, 
and the assumption of the theorem that \(\cov(\hat{Y}_{X,Z}, Y \mid X) \leq 0\) to establish the inequality. 
The above shows that the second summand on the right-hand side of Eq.~\eqref{eq:diff} is non-negative. Since we previously showed that the first summand is non-negative (and strictly positive under the additional assumption), this proves the result.

\(\qed\)

\section{Proof of Corollary~\ref{cor:main}}

By Theorem~\ref{thm:main}, it is sufficient to show that \(\cov(\hat{Y}_{X,Z}, Y \mid X) \leq 0\).
We start by noting that
\begin{align*}
\cov(\hat{Y}_{X,Z}, Y \mid X) &= \cov(f(X) + g(X) \cdot Z, Y \mid X) \\
&= g(X) \cdot \cov(Y, Z \mid X),
\end{align*}
and so, by the assumption of the theorem,
\begin{equation}
\label{eq:sign}
\sign\left(\cov\left(\hat{Y}_{X,Z}, Y \mid X\right)\right) =
-\sign\left(g(X) \cdot \cov\left(Y', Z \mid X\right)\right).
\end{equation}
Now, by repeatedly applying the law of iterated expectations, we have
\begin{align*}
\EE\left[Z\cdot Y' \mid X\right] &= \EE\left[\EE\left[Z\cdot Y' \mid X,Z\right]\mid X\right]\\
&= \EE\left[Z \cdot \EE\left[Y' \mid X,Z\right]\mid X\right]\\
&= \EE\left[Z \cdot \hat{Y}_{X,Z} \mid X\right]\\
&= f(X) \cdot \EE[Z \mid X] + g(X) \cdot \EE\left[Z^2 \mid X\right].
\end{align*}
Similarly, we have
\begin{align*}
\EE[Y' \mid X] &= \EE[ \EE[Y' \mid X, Z] \mid X]\\
&= \EE[\hat{Y}_{X,Z} \mid X]\\
&= f(X) + g(X) \cdot \EE[Z \mid X].
\end{align*}
Putting the above together, we get
\begin{align*}
\cov\left(Y', Z \mid X\right) &= \EE\left[Z\cdot Y' \mid X\right] - \EE[Y' \mid X] \cdot \EE[Z \mid X]\\
&= g(X)\left(\EE\left[Z^2 \mid X\right] - \EE\left[Z \mid X\right]^2\right)\\
&= g(X)\cdot \var(Z \mid X).
\end{align*}
Finally, by Eq.~\eqref{eq:sign},
\begin{align*}
\sign\left(\cov\left(\hat{Y}_{X,Z}, Y \mid X\right)\right) &=
-\sign\left(g(X)^2 \cdot \var(Z \mid X)\right)\\
&\leq 0,
\end{align*}
establishing the result.

\(\qed\)

\section{Kitchen-Sink Models and Independent Noise}

When the proxy label \(Y'\) and the true label \(Y\) simply differ by additive, independent noise, then it is advantageous to use all available information when constructing risk scores. The following proposition formalizes this statement.

\begin{proposition}
\label{prop:noise}
    In the setting of Theorem~\ref{thm:main}, suppose \(Y' = Y + S\) where \(S \indep X,Z \). Then 
    \begin{equation*}
    \EE\left[\left(\hat{Y}_{X,Z} - Y\right)^2\right] \leq \EE\left[\left(\hat{Y}_X - Y\right)^2\right]. 
    \end{equation*}
\end{proposition}

\noindent \emph{Proof.}
First note that
\begin{align*}
    \hat{Y}_{X,Z} &= \EE[Y \mid X,Z] + \EE[S \mid X,Z] \\
    &= \EE[Y \mid X,Z] + \EE[S],
\end{align*}
where the second equality uses the independence assumption. Similarly,
\begin{align*}
    \hat{Y}_{X} &= \EE[Y \mid X] + \EE[S \mid X] \\
    &= \EE[Y \mid X] + \EE[S].
\end{align*}
Now, using the notation \(Y_{X,Z} = \EE[Y \mid X,Z]\) and 
\(Y_X = \EE[Y \mid X]\), we have
\begin{align*}
    \EE\left[\left(\hat{Y}_{X,Z} - Y\right)^2\right] 
    - \EE\left[\left(\hat{Y}_{X} - Y\right)^2\right] \\
    &\hspace{-2in}= \EE\left[\big(Y_{X,Z} - Y + \EE[S]\big)^2\right]
    - \EE\left[\big(Y_{X} - Y + \EE[S]\big)^2\right] \\
    &\hspace{-2in}= \EE\left[\big(Y_{X,Z} - Y\big)^2\right] - \EE\left[\big(Y_{X} - Y\big)^2\right] + 2\EE[S]\big(\EE[Y_{X,Z}-Y] - \EE[Y_{X}-Y]\big)\\
    &\hspace{-2in}= \EE\left[\big(Y_{X,Z} - Y\big)^2\right] - \EE\left[\big(Y_{X} - Y\big)^2\right]\\
    &\hspace{-2in}= \EE \left [\EE\left[\big(Y_{X,Z} - Y\big)^2 \mid X,Z\right] \right]
    - \EE \left [\EE\left[\big(Y_{X} - Y\big)^2 \mid X,Z \right]\right],
\end{align*}
where the third equality follows from the fact that 
\(\EE[Y_{X,Z}] = \EE[Y_{X}] = \EE[Y]\),
and the last equality follows from the law of iterated expectations.
Finally, since 
\begin{equation*}
\arg \min_{c}  \EE\left[\big(c - Y\big)^2 \mid X,Z \right] = Y_{X,Z},
\end{equation*}
we have that
\begin{equation*}
    \EE\left[\big(Y_{X,Z} - Y\big)^2 \mid X,Z\right]
    - \EE\left[\big(Y_{X} - Y\big)^2 \mid X,Z \right] \leq 0,
\end{equation*}
establishing the result.

\(\qed\)

\section{A Stylized Model of Arrest and Behavior}

We formally describe and analyze the SEM depicted in Figure~\ref{fig:sim-dag}. Our model has three independent exogenous variables 
\(U_Z = \N(0, \sigma_Z^2)\), 
\(U_{A_0} = \N(0, \sigma_A^2)\),
and \(U_{A_1} = \N(0, \sigma_A^2)\).
We additionally have two correlated
exogenous variables
\(U_{B_0} = \N(0, \sigma_B^2)\)
and \(U_{B_1} = \N(0, \sigma_B^2)\)
that are independent of the first three, 
with \(\cov(U_{B_0}, U_{B_1}) = \delta \geq 0\).
Now, for non-negative constants \(\alpha\), \(\beta\), and \(\gamma\), the key variables in the model are generated by the following linear structural equations:
\begin{align}
\begin{split}
  Z &= U_Z, \\
  B_0 &= \beta Z + U_{B_0}, \\
  B_1 &= \beta Z + U_{B_1}, \\
  A_0 &= \alpha Z + \gamma B_0 + U_{A_0}, \\
  A_1 &= \alpha Z + \gamma B_1 + U_{A_1}. \\
\end{split}
\end{align}
We set the variances of the exogenous variables 
(\(\sigma_Z^2\), \(\sigma_A^2\), and \(\sigma_B^2\))
in a manner that ensures that the remaining variables 
(\(Z\), \(B_0\), \(B_1\), \(A_0\), and \(A_1\))
are standardized, meaning they have mean 0 and variance 1---we show how to do this below.
We can thus interpret their values as representing the extent to which individuals differ from the population averages.
In the case of neighborhood (\(Z\)), we can think of its value as denoting the level of police enforcement in an area.

To start, we set \(\sigma_Z^2 = 1\), which ensures \(\var(Z) = 1\).
Now, since \(Z \indep U_{B_0}\), we have that \(\var(B_0) = \beta^2 + \sigma_B^2\).
Consequently, setting \(\sigma_B^2 = 1 - \beta^2\) ensures that
\(\var(B_0) = 1\) (and, similarly, that \(\var(B_1) = 1\)). Finally,
as above,
\(\var(A_0) = \alpha^2 + \gamma^2 + \sigma_A^2 + 2\alpha \gamma\cov(Z,B_0)\).
One especially nice aspect of linear graphical models is that the covariance between any two variables can be immediately computed from the edge weights via the
the Wright rules~\cite{wright1921systems,pearl2013linear}.
Specifically, when the nodes are standardized to have variance 1, then the covariance between any two variables in the graph is the sum, over all \(d\)-connected paths between the variables, of the product of the edge weights along the path.
A path is \(d\)-connected if it does not pass through any colliders (i.e., nodes with head-to-head arrows along the path).
To compute \(\cov(Z,B_0)\),
observe that the only \(d\)-connected path between \(Z\)
and \(B_0\) is the direct path from \(Z\) to \(B_0\), having edge weight \(\beta\).
As a result, \(\cov(Z,B_0) = \beta\),
meaning that setting
\(\sigma_A^2 = 1 - \alpha^2 - \gamma^2 - 2 \alpha\beta\gamma\)
ensures that \(A_0\) (and, analogously, \(A_1\)) have unit variance. Recapping, we have
\begin{align}
\label{eq:u-var}
\begin{split}
  \sigma_Z^2 &= 1,\\
  \sigma_B^2 &= 1-\beta^2, \\
  \sigma_A^2 &= 1 - \alpha^2 - \gamma^2 - 2 \alpha\beta \gamma.
\end{split}
\end{align}
Our model is thus described by the four non-negative parameters 
\(\alpha\), \(\beta\), \(\gamma\), and \(\delta\), depicted as edge weights in 
Figure~\ref{fig:sim-dag},
with the constraint that the quantities in 
Eq.~\eqref{eq:u-var} are non-negative.
Those constraints in turn imply that the parameters are each less than or equal to 1.

Our theoretical results in Theorem~\ref{thm:main} and 
Corollary~\ref{cor:main} require understanding the conditional distributions of model features. 
For multivariate normal random variables, these conditional distributions can be computed analytically~\cite{eaton1983multivariate}, allowing us to examine properties of our motivating SEM in more depth.
Specifically, suppose \(\mathbf{W}\) is a \(k\)-dimensional multivariate normal random variable with mean \(\boldsymbol{\mu}\) and covariance \(\mathbf{\Sigma}\), which we partition
into into its first \(q\) components and its remaining \(k-q\) components: \(\mathbf{W} = [\mathbf{W_1}, \mathbf{W_2}\)].
Further suppose we accordingly partition \(\boldsymbol{\mu}\) and \(\mathbf{\Sigma}\) into its components:
\begin{align*}
\boldsymbol{\mu} &= 
\begin{bmatrix}
\boldsymbol{\mu_1} \\
\boldsymbol{\mu_2}
\end{bmatrix}
\ \text{with sizes} 
\begin{bmatrix}
q \times 1 \\
(k-q) \times 1
\end{bmatrix},\\
\boldsymbol{\Sigma} &= 
\begin{bmatrix}
\mathbf{\Sigma_{11}} & \mathbf{\Sigma_{12}}\\
\mathbf{\Sigma_{21}} & \mathbf{\Sigma_{22}}
\end{bmatrix}
\ \text{with sizes} 
\begin{bmatrix}
q \times q & q \times (k-q)\\
(k-q) \times q & (k-q) \times (k-q)
\end{bmatrix}.
\end{align*}
Then the distribution of \(\mathbf{W_1}\) conditional on 
\(\mathbf{W_2}\) is multivariate normal with 
mean 
\[\boldsymbol{\mu_1} + \mathbf{\Sigma_{12}} \mathbf{\Sigma_{22}}^{-1}(\mathbf{W_2} - \boldsymbol{\mu_2})\]
and covariance
\[\mathbf{\Sigma_{11}} - \mathbf{\Sigma_{12}} \mathbf{\Sigma_{22}}^{-1} \mathbf{\Sigma_{21}}.\]

As a result, the linearity assumption of Corollary~\ref{cor:main} is satisfied for multivariate normal random variables. 
In particular, in our motivating example, the conditional distribution of \(A_1\) 
given \(A_0\) and \(Z\) is normal, with
\begin{align*}
\EE[A_1 \mid A_0, Z] &=
\begin{bmatrix}
\sigma_{A_1 A_0} & \sigma_{A_1 Z} 
\end{bmatrix} 
\begin{bmatrix}
1 & \sigma_{A_0 Z} \\
\sigma_{A_0 Z} & 1 
\end{bmatrix}^{-1}
\begin{bmatrix}
A_0 \\
Z
\end{bmatrix} \\
&=
\frac{1}{1-\sigma_{A_0 Z}^2}
\begin{bmatrix}
\sigma_{A_1 A_0} & \sigma_{A_1 Z} 
\end{bmatrix} 
\begin{bmatrix}
1 & -\sigma_{A_0 Z} \\
-\sigma_{A_0 Z} & 1 
\end{bmatrix}
\begin{bmatrix}
A_0 \\
Z
\end{bmatrix}\\
&=
\frac{\sigma_{A_1 A_0} - \sigma_{A_1 Z} \cdot \sigma_{A_0 Z}}{1-\sigma_{A_0 Z}^2} A_0 
+ \frac{\sigma_{A_1 Z} - \sigma_{A_1 A_0} \cdot \sigma_{A_0 Z}}{1-\sigma_{A_0 Z}^2} Z,
\end{align*}
where the \(\sigma\) notation denotes the covariance of the indexed random variables.

Further,
the conditional distribution of \((A_1, Z)\) given \(A_0\) is likewise multivariate normal, with covariance matrix
\begin{align*}
\begin{bmatrix}
1 & \sigma_{A_1 Z} \\
\sigma_{A_1 Z} & 1 
\end{bmatrix} - 
\begin{bmatrix}
\sigma_{A_1 A_0} \\
\sigma_{A_0 Z}
\end{bmatrix}
\begin{bmatrix}
\sigma_{A_1 A_0} & \sigma_{A_0 Z}
\end{bmatrix} 
&= 
\begin{bmatrix}
1 & \sigma_{A_1 Z} \\
\sigma_{A_1 Z} & 1 
\end{bmatrix} - 
\begin{bmatrix}
\sigma_{A_1 A_0}^2 & \sigma_{A_1 A_0} \cdot \sigma_{A_0 Z} \\
\sigma_{A_1 A_0} \cdot \sigma_{A_0 Z} & \sigma_{A_0 Z}^2 
\end{bmatrix}\\
&= \begin{bmatrix}
1-\sigma_{A_1 A_0}^2 & \sigma_{A_1 Z} - \sigma_{A_1 A_0} \cdot \sigma_{A_0 Z} \\
\sigma_{A_1 Z} - \sigma_{A_1 A_0} \cdot \sigma_{A_0 Z} & 1-\sigma_{A_0 Z}^2 
\end{bmatrix}.
\end{align*}
Consequently, 
\begin{equation}
\label{eq:cov_A1_Z_A0}
\cov(A_1, Z \mid A_0) = \sigma_{A_1 Z} - \sigma_{A_1 A_0} \cdot \sigma_{A_0 Z},
\end{equation}
and, analogously, we have that
\begin{equation}
\label{eq:eq:cov_B1_Z_A0}
\cov(B_1, Z \mid A_0) = \sigma_{B_1 Z} - \sigma_{B_1 A_0} \cdot \sigma_{A_0 Z}.
\end{equation}

As above, we can compute the covariances in Eqs.~\eqref{eq:cov_A1_Z_A0} and \eqref{eq:eq:cov_B1_Z_A0} 
via the Wright rules.
For example, as seen in Figure~\ref{fig:sim-dag}, there are two \(d\)-connected paths between \(Z\) and \(A_0\):
the direct connection with edge weight \(\alpha\);
and the path through \(B_0\), with product of edge weights \(\beta \gamma\).
Consequently, \(\cov(Z, A_0) = \alpha + \beta \gamma\).
This procedure allows us to compute all of the terms appearing on the right-hand side of  Eqs.~\eqref{eq:cov_A1_Z_A0} and \eqref{eq:eq:cov_B1_Z_A0}, yielding:
\begin{align}
\label{eq:pairwise-cov}
\begin{split}
\sigma_{A_0 Z} &= \alpha + \beta \gamma \\
\sigma_{A_1 Z} &= \alpha + \beta \gamma \\
\sigma_{B_1 Z} &= \beta \\
\sigma_{A_1 A_0} &= \alpha^2 + 2\alpha\beta\gamma + \beta^2\gamma^2 + \gamma^2\delta\\
\sigma_{B_1 A_0} &= \alpha\beta + \beta^2\gamma + \gamma\delta.
\end{split}
\end{align}

Leveraging the above, we now show that 
\(\cov(A_1, Z \mid A_0) \geq 0\), meaning that neighborhood is positively correlated with future arrests, conditional on past arrests.
To see this, first note that
\begin{align*}
    \delta &= \cov(U_{B_0}, U_{B_1}) \\
    &\leq \sigma_B^2\\
    &= 1 - \beta^2,
\end{align*}
and so \(\beta^2 + \delta \leq 1\). Now,
\begin{align*}
    \cov(A_1, Z \mid A_0) 
    &= \sigma_{A_1 Z} - \sigma_{A_1 A_0} \cdot \sigma_{A_0 Z}\\
    &= \alpha + \beta \gamma - (\alpha + \beta \gamma) \cdot (\alpha^2 + 2\alpha\beta\gamma + \beta^2\gamma^2 + \gamma^2\delta)\\
    &=(\alpha + \beta \gamma)\cdot(1 - \alpha^2 - 2\alpha\beta\gamma - \beta^2\gamma^2 - \gamma^2\delta)\\
    &=(\alpha + \beta \gamma)\cdot(1 - \alpha^2 - 2\alpha\beta\gamma - \gamma^2(\beta^2 +  \delta))\\
    &\geq (\alpha + \beta \gamma)\cdot(1 - \alpha^2 - 2\alpha\beta\gamma - \gamma^2)\\
    &=  (\alpha + \beta \gamma) \cdot \sigma_A^2\\
    &\geq 0,
\end{align*}
where the first inequality follows from the fact that \(\beta^2 + \delta \leq 1\).

Next we consider \(\cov(B_1, Z \mid A_0)\), and note that
\begin{align*}
    \cov(B_1, Z \mid A_0) 
    &= \sigma_{B_1 Z} - \sigma_{B_1 A_0} \cdot \sigma_{A_0 Z} \\
    &= \beta - (\alpha\beta + \beta^2\gamma + \gamma\delta)\cdot(\alpha + \beta\gamma).
\end{align*}
In particular, when \(\beta = 0\), 
meaning that neighborhood does not impact behavior,
then 
\begin{equation*}
\cov(B_1, Z \mid A_0) = -\alpha \gamma \delta.
\end{equation*}
In other words, when neighborhood does not impact behavior (i.e., when \(\beta = 0\)),
neighborhood is negatively correlated with future behavior 
conditional on past arrests.
(And, by the above, neighborhood is always positively correlated with future arrests conditional on past arrests.)
By Corollary~\ref{cor:main}, it is thus better in this case to base predictions of future behavior solely on past arrests, excluding neighborhood,
as we see in Figure~\ref{fig:full-sim}.

\end{document}